\documentclass[10pt]{article}

\usepackage{url}      
\usepackage{booktabs} 
\usepackage{colortbl} 
\usepackage{xcolor}   
\usepackage{tikz}     
\usepackage{wasysym}  
\usepackage{tabularx} 
\usepackage[hidelinks]{hyperref} 

\newcommand{\rcolor}{\rowcolor{gray!20}} 

\newcommand{\pie}[1]{
\begin{tikzpicture}
 \draw (0,0) circle (1ex);\fill (1ex,0) arc (0:#1:1ex) -- (0,0) -- cycle;
\end{tikzpicture}
}

\newcommand{\spie}[1]{ 
\begin{tikzpicture}
 \draw (0,0) circle (0.75ex);\fill (0.75ex,0) arc (0:#1:0.75ex) -- (0,0) -- cycle;
\end{tikzpicture}
}

\newcommand{\p}[1]{$P_{#1}$}

\newcommand{\q}[1]{{\small \sf ``#1''}}

\newcommand{\qq}[2]{{\small \sf ``#1''~({\small{$P_{#2}$}})}}

\newcommand{\mc}[1]{{\small\textsf{#1}}}

\begin{document}

\title{How Users Consider Web Tracking\\When Seeking Health Information Online}
\author{Martin P. Robillard, Linh V. Nguyen, Deeksha M. Arya, and Jin L.C. Guo\vspace{2mm}\\
{\normalsize McGill University, Montréal, QC, Canada}\\
{\small robillard@acm.org ORCID: 0000-0002-0248-1384}\\
{\small viet.l.nguyen@mail.mcgill.ca ORCID: 0009-0006-7835-8442}\\
{\small deeksha.arya@mail.mcgill.ca ORCID: 0000-0002-3719-5011}\\
{\small jguo@cs.mcgill.ca ORCID: 0000-0003-1782-1545}}

\date{26 March 2026\vspace{5mm}}
\maketitle

\thispagestyle{empty} 

\begin{abstract}
Health information websites offer instantaneous access to information, but have important privacy implications as they can associate a visitor with specific medical conditions. We interviewed 35 residents of Canada to better understand whether and how online health information seekers exercise three potential means of protection against surveillance: website selection, privacy-enhancing technologies, and self-censorship, as well as their understanding of web tracking. Our findings reveal how users' limited initiative and effectiveness in protecting their privacy could be associated with a missing or inaccurate understanding of how implicit data collection by third parties takes place on the web, and who collects the data. We conclude that to help health information seekers better protect their online privacy, we may need to shift privacy awareness efforts from what information is collected to how it is collected.
\end{abstract}

\section{Introduction}
\label{s:introduction}

Online health information seeking is prevalent~\cite{Jia2021, Rains2018}. 
People seek health information online for a variety of reasons that include learning about one's health condition, increasing one's sense of control over one's health, and learning from the experience of others who share a health condition~\cite{BootMeijman2010}. There are different ways to access health information online. One common way is through a \textit{health information website}: a website that presents health information in the form of a \textit{structured} collection of mostly \textit{text documents} on \textit{health topics} intended for the \textit{general public}. Other activities can include using social media~\cite{DeChoudhury2014}, AI chat bots~\cite{LundEtAl2025}, or accessing web resources that are not textual or part of information websites (e.g., videos~\cite{LiuZhang2024a}). 

Seeking health information online has important privacy implications. 
Websites can not only collect data directly for their operator, they also \textit{leak} information about users to third parties such as advertisers or social media companies~\cite{Libert2015}. 
For example, when a user visits a webpage associated with a medical condition, this page can contain requests to load third-party resources such as analytics code or social media ``like'' buttons~\cite{CG2019a}. 
Through such mechanisms, third parties can discover that a user made a request for information on a medical condition associated with the visited webpage and use information collected from the networking interaction to link the request with the user~\cite{Libert2015,OPC2014,New2022a,ZWE2025a}. The consequent threats include persistent and unsettling ads~\cite{ULP2012a}, differential pricing~\cite{HannakSLMW14}, and algorithmic discrimination~\cite{custers2013data}.

The goal of our research is to help Internet users achieve better protection against third-party tracking when visiting health information websites. Towards that end, we sought to better understand whether and
how health information seekers exercise three potential means of protection against surveillance and how they understand how their personal data is collected when they browse the web. The three means of data protection we investigated are: 1) discriminating between different health information websites (e.g., commercial vs. governmental), 2) using privacy-enhancing technology, and 3) self-censoring their searches. 
We investigated health information seeker's privacy considerations through qualitative interviews with 35 adults residents of Canada with a diversity of characteristics including gender, age group, educational background, digital literacy, and privacy attitude.

Consistent with recent work~\cite{Boerman2021, Kyi2024}, we observed that our study participants' efforts at protecting their privacy were limited. In addition, our interactions with participants surfaced a key aspect of their detachment from privacy concerns, namely a limited understanding of the process by which their browsing activities are leaked to third parties. When discussing different privacy-related considerations, participants used the passive voice or employed generic abstractions to refer to data recipients. This observation helps explain important themes in privacy research, such as the lack of motivation to protect one's privacy. While previous research focused on eliciting Internet users' attitudes and tactics related to data protection, less attention was devolved to the  details of how users think about data collection, in particular in the context of seeking health information online. Our study addresses this gap.

Using qualitative methods, this study provides a detailed view of how individuals seek technical health information on websites, in the Canadian context. It exposes a parallel between the limited effort they invest in protecting their browsing history and their partial understanding of the tracking technology that enables the surveillance economy. The findings of this study can inform the decisions of public website operators, help users protect themselves by choosing websites with better privacy practices, and motivate the design of educational material that raises awareness about online surveillance.

\section{Background}
\label{s:Background}

While seeking health information, the browsing activities of Internet users are monitored by various organizations that may not have users' best interests  at heart. In Section~\ref{ss:WebTracking}, we explain how this \textit{web tracking} takes place and the major technical findings that relate to it. We complement this overview with an analysis of the level of user tracking deployed on popular health information websites (Section~\ref{ss:HealthPortals}).

\subsection{Web Tracking and Surveillance}
\label{ss:WebTracking}

There is much evidence showing that technologies are being used to collect, store, analyze, and share users' information and behavior when they browse online health resources (e.g.,~\cite{Huesch2013, New2022a, Libert2015}). One of the most widespread privacy threats when browsing online stems from \textit{third-party tracking}~\cite{CG2019a}. When users engage with an online service, for example Facebook, this service is the \textit{first party}. Users (the second party), are usually conscious that they are disclosing personal information to first parties and others have studied this type of information disclosure extensively (e.g.,~\cite{Anarky2021, Surariu2025}). Current web technology, however, enables organizations other than the first party to collect web browsing information~\cite{Libert2015-1}. This leak of personal information is enabled when the first-party website loads resources such as social media buttons or analytics libraries, which connect to a \textit{third-party} server. An important issue with third-party tracking is that personal information is collected about users \textit{implicitly}, without the user having to take any action~\cite{Roesner2012}. Consequently, users can easily remain unaware that their browsing history is shared with third parties.

We exemplify this threat with a simplified scenario. In a web browsing session, we can navigate to the website of the World Health Organization and choose the page on \textit{Abortion}. The web browser loads the page, which is organized into different sections that can be accessed via mouse clicks. At the same time, the page loads technology that contacts servers belonging to Microsoft, Google, and Facebook, along with a commercial fundraising platform and other website optimization services.\footnote{\url{https://www.who.int/health-topics/abortion}. Accessed 2025-12-16 from a Canadian location. Tracking information reported by Privacy Badger~\cite{PrivacyBadger}. Results may vary based on connection settings.}

The connection to a third-party server via a web page leaks various types of personal information, such as a user's IP address. Additional technology deployed either on the first-party website or on the third-party server enables the collection of a wealth of additional information that can personally identify a user. This information can then be aggregated with existing information about the user to form a detailed personal profile~\cite{Libert2015}.

Web tracking technology and techniques are continually evolving in a type of arms race to stay ahead of technical defenses and privacy protection legislation~\cite{BCV2017a}. Tracking also involves an element of social engineering. For example, websites can provide a perception of organizations' ethical care for privacy that encourages users to disclose accurate, personal information~\cite{Thompson2022}. Website design can also mislead users to collect their information implicitly and explicitly~\cite{Papadogiannakis2021}. Finally, how third parties \textit{use} web visitor data is usually proprietary and confidential information. However, initial evidence shows that the data is used to influence user behavior through advertisement~\cite{ZWE2025a} and dynamic pricing, also known as \textit{surveillance pricing}~\cite{FTC2025}. The consequences of web tracking are also not limited to ads, and can include the realization of any number of privacy threats~\cite{Solove2006}.

\subsection{Tracking on Health Websites}
\label{ss:HealthPortals}

We illustrate the current context of web tracking on health information websites by reporting on relevant websites, the number of articles they offer, and the web tracking deployed on them.

In Canada, in addition to the federal government, all ten provinces and one of the three territories provide a public health information website. 
We complemented this list by considering other popular health websites likely to be accessed by English speakers in general. 
For this purpose, we accessed the SimilarWeb service and obtained the 50 most popular websites worldwide in the \textit{Health} category according to their proprietary ranking algorithm, and filtered the list according to specific criteria (see Appendix~\ref{a:websites} for details).
With this process, we identified an additional ten websites: six commercial health websites, two US government websites (CDC and MedlinePlus), one UK government website (National Health Service, NHS), and the website of the World Health Organization (WHO).

The websites in our dataset vary in the amount of information they offer. Some only provide a few articles on key public health topics, while others provide encyclopedic coverage with thousands of articles. For each website, we semi-automatically identified the articles which are related to a health concern, i.e. about a symptom, condition, or treatment related to a person's health. Appendix~\ref{a:urls} provides the Uniform Resource Locator (URL) for each website along with the details of how we analyzed them.

\begin{table}[ht!]
    \centering
        \caption{Content and surveillance metrics for selected health information websites, in order of decreasing number of articles. The column \textit{Articles} provides the number of applicable articles on the website (M=767, SD=888). In the column \emph{3rd Parties}, we report the number of different organizations who own a third-party domain reached via tracking technology deployed on the site (M=21.5, SD=31.7). As Google is the dominant third party contacted by websites~\cite{Libert2015}, we also indicate if it owns a domain contacted by the site.
        Grayed rows indicate \emph{commercial websites}, which include non-profit organizations that carry out commercial activities. 
        }
    \begin{tabular}{lrrc}
    \toprule
    \textbf{Website}  &  \textbf{Articles} & \textbf{3rd Parties} & \textbf{Google?}\\
    \midrule
    \rcolor Cleveland Clinic          & 3653 &  55 & $\bullet$\\
            Alberta                   & 2031 &   1 & $\bullet$\\
    \rcolor WebMD                     & 1609 &  80 & $\bullet$\\
            CDC                       & 1580 &   2 & $\bullet$\\
            Ontario                   & 1381 &   0 & \\
            NHS                       & 1202 &   0 & \\
    \rcolor Mayo Clinic               & 1152 &  60 & $\bullet$\\
            British Columbia          & 1140 &   1 & \\
            Medline Plus              & 1023 &   4 & $\bullet$\\
    \rcolor Everyday Health           & 618  &  72 & $\bullet$\\
    \rcolor Verywell Health           & 326  & 105 & $\bullet$\\
            Saskatchewan              & 293  &  25 & $\bullet$\\
            Canada                    & 242  &   2 & \\
            WHO                       & 192  &   3 & $\bullet$\\
    \rcolor Healthline                & 150  &  48 & $\bullet$\\
            Qu\'{e}bec                & 90   &   2 & $\bullet$\\
            Manitoba                  & 78   &   5 & $\bullet$\\
            Nova Scotia               & 31   &   2 & $\bullet$\\
            Newfoundland \& Labrador  & 26   &   2 & $\bullet$\\
            New Brunswick             & 25   &   1 & $\bullet$\\
            Northwest Territories     & 22   &   1 & $\bullet$\\
            Prince Edward Island      & 14   &   3 & $\bullet$\\
    \bottomrule
    \end{tabular}
    \label{t:PortalContent}
    \end{table}

We assessed the nature and extent of user tracking on each website using the \textit{Blacklight privacy inspector}. Blacklight is web application that automatically visits a website to monitor ``which scripts on that website are potentially surveilling the user by performing seven different tests, each investigating a specific, known method of surveillance''~\cite{blacklightBlog}. For each website, we executed the tool using as input an arbitrary article page as a representative. 
We ran the tests in May 2024.
We collected the domain of all trackers as reported by Blacklight, then joined this data with DuckDuckGo's Tracker Radar database~\cite{trackerRadar2020} to obtain the owner of each domain, excluding the website's own domain.\footnote{Assessing the amount of tracking done by a website is an approximate process due to the diversity of the technology involved and the obscurity of some of the practices. The developers of the Blacklight tool acknowledge several limitations related to the heuristic nature of the techniques they use to detect tracking. We reiterate Blacklight's disclaimer that these results ``should not be taken as the final word on potential privacy violations by a given website''~\cite{blacklightBlog}.}

Table~\ref{t:PortalContent} describes the tracking on selected health information websites. 
Our data shows that extensive tracking takes place on some health information websites, and that commercial websites comparatively track users extensively and disseminate the information collected to a large group of third parties. Almost all websites in our sample send data to Google, the organization that has also been reported to conduct the most web tracking on popularly visited health websites~\cite{Libert2015}. Our analysis also reveals that some governmental health websites, such as that of Saskatchewan and Manitoba, also deploy commercial surveillance. Thus, users are justified in their concern of online privacy despite governmental regulations~\cite{Prince2021}. However, some websites do exhibit good privacy practices: the UK's National Health Service and the Province of Ontario both have health websites that do not appear to send any data to third parties. 
\section{Literature Review}
\label{s:Literature}

We begin by describing how Internet users seek information online and leverage this literature to scope the present work. We then discuss the theoretical foundations of research on privacy that the present work is built on, including privacy calculus, protection motivation theory, and contextual integrity. Finally, we review key findings about the extent to which users are aware of these privacy invasions, and what they do to protect themselves.

\subsection{Seeking Health Information Online}
\label{ss:SeekingHealthInformation}

Searches for health-related information can be compared to general information seeking behavior~\cite{Jia2021, Kim1015}. Numerous factors can influence how people search for information online. In their survey of 41 eye-tracking studies of how users select web search results~\cite{LK2021a}, Lewandowski and Kammerer report on the potential effect of the position of a result in a list, the role of domain knowledge, and various demographic factors. Their conclusion includes that ``both grade level and prior topic knowledge seem to be related to a more systematic and more careful reading of the [results] in order to decide which information sources to access''~\cite[p.1505]{LK2021a}. However, none of the studies reviewed in their work examined users' motivation to protect their privacy, and the investigation of such motivational factors is highlighted as an area for future research. Our present work addresses this gap. In the remainder of this section, we review the literature on specific factors that we hypothesize could be modulated by the presence of web tracking.

\paragraph{Object of the Information Search} An important difference exists between searching for \textit{technical} vs. \textit{experiential} health information. According to Rains (2018), ``technical information involves the medical  aspects of a health condition, whereas experiential information focuses on  the personal experiences of people coping with illness'', with technical information  ``more likely to appear on health websites such as those operated by the Centers for Disease Control...''~\cite[p.91]{Rains2018}. In contrast, experiential information can be gleaned more easily from social media services such as Facebook or Reddit. As our work targets the use of health information websites, we focus on the search for technical information.

\paragraph{Trust}
The concept of trust is pivotal in research on web search and it has been studied extensively, including specifically for seeking heath information~\cite{Kim2016}. People consider \textit{trust} in a website as a factor when interacting with it~\cite{Martin2016}. For example, whether an individual trusts the content or the people behind a website impacts their decision to reveal personal information to the website~\cite{Dinev2006, Meier2024} and even whether to authorize sharing of their data with third parties~\cite{Libaque-Saenz2016}. Thus, an individual's perception of whether a website is credible and trustworthy can play an important role in how users choose to disclose information. The concept of trust, however, is multifaceted. Two facets that are especially relevant to seeking technical health information are \textit{trust in the website's information quality}~\cite{Jin2025}, vs. \textit{trust that the website respects users' privacy}~\cite{Knowles2023}.  

\paragraph{Stigma}
Some health conditions are surrounded by stigma, and this aspects can impact the way people browse online and share details about their conditions. Based on a comparison of over 7000 survey respondents with an illness categorized as either stigmatized or non-stigmatized, Berger et al. found partial support for the hypothesis that people with a stigmatized illness were \textit{more} likely to use the Internet to seek health information~\cite{Berger_Wagner_Baker_2005}. This outcome, however, was observed before the onset of mass commercial surveillance and increased awareness of online privacy concerns. It is now worth considering how health information seekers think of the issue in the presence of pervasive web tracking. In a more recent study, De Choudhury et al. investigated specifically whether participant's health-related online activities are impacted by ``stigma associated with the health condition''~\cite{DeChoudhury2014}. Through a survey and the analysis of web log data, they concluded that, for conditions with high stigma, there is more evidence of activity using search engines than on Twitter. This observation supports the hypothesis that users are indeed sensitive to the stigmatizing nature of the information they disclose on the Internet. Still, despite risks to privacy, people prefer to search for health information online because it provides a sense of privacy in terms of being able to interact with online communities with apparent confidentiality~\cite{Jia2021}. 

\subsection{Theoretical Foundations}
\label{ss:TheoriesOfPrivacy}

In this work, we construe privacy from the perspective of information disclosure~\cite{knijnenburgModernSocioTechnicalPerspectives2022, Suddekunte2024}. Specifically, we examine Internet users' awareness and attitudes toward data collection and use by other parties and how this reasoning process modulates potential attempts to control voluntary or involuntary disclosure. In particular, our work is informed by the logic of \textit{privacy calculus} and the \textit{protection motivation} and \textit{contextual integrity theories}.

\paragraph{Privacy Calculus}
Privacy calculus~\cite{culnan1999information, Meier2024} frames the disclosure of personal information as a cost-benefit analysis. This conceptualization aligns with the typification of ``privacy pragmatists'' in Westin's privacy taxonomy~\cite{Westin2003}, where users examine the benefits and risks of the data collection and use before``decid[ing] whether to trust the organization or seek legal oversight''. However, this theory has been criticized for its assumption that users are rational~\cite{acquistiDigitalPrivacyTheory2007} and its limited ability to explain the \textit{privacy paradox}, where, despite acknowledging privacy risks, users still forego privacy protection options. Another limitation of the privacy calculus theory is that it offers few insights for situations wherein the collection of personal information is implicit and likely occurs without the knowledge and consent of users. Several improvements to privacy calculus have been proposed, mainly by integrating other theoretical perspectives, e.g., factoring in social values~\cite{Tsirozidis2025} or users' demographic information~\cite{epsteinDimensionalisedPrivacyBehaviour2025}. However, this prior work primarily considers users' \textit{intentional} information disclosure, particularly to a single party. In this study, we focus on implicit data collection.

\paragraph{Protection Motivation Theory} 
Protection motivation theory (PMT) models how individuals adopt protective behavior in response to appeals to sentiments of fear~\cite{Rogers1975, Maddux1983}. The PMT surfaces the association between individual's protective motivation in terms of  two main factors: \textit{threat appraisal} and \textit{coping appraisal}. Compared with privacy calculus, which focuses on risk-benefit analysis, PMT places greater emphasis on the interplay between the perceived significance of threats and the individual's perceived ability to mitigate them through effective responses. PMT has been applied in the context of privacy by, e.g., Boerman et al.~\cite{Boerman2021}, who report that ``people rarely to occasionally protect their online privacy'' and that people have low confidence in their ability to protect against threats to their privacy. Through a survey study, Chennamaneni and Gupta~\cite{Chennamaneni2023} found that threat appraisal has a significant effect on model of users' privacy concerns. Self-efficacy also has a significant effect on users' intention to engage in privacy protection behaviors. Our study complements this work by providing a qualitative overview of how users perceive the threats related to seeking health information online and how they might engage with privacy-enhanced technologies during this process.

\paragraph{Contextual Integrity}
The theory of contextual integrity treats privacy as situational and governed by norms~\cite{Nissenbaum2010}. Within this framework, privacy is considered maintained when both the types of information and how it flows align with established expectations of appropriateness. Conversely, a violation occurs when data practices deviate from these context-specific norms. Contextual integrity affords the examination of how culture and value shape online behaviors, such as photo sharing~\cite{kumarModernDayBaby2015}. It can also be leveraged to improve online privacy literacy by explicitly educating and reflecting on the information type, subject, sender and receiver of the information flow during online activities~\cite{Suddekunte2024}. The present study contributes to the qualification of the norm for online health information seeking: the nature of the search queries, how sensitive the searches are perceived by the users, and to what extent users are aware of the implicit information flow to third parties.

\subsection{Privacy Awareness and Protection}
\label{ss:PrivacyAwareness}

We assume that understanding the threats is a necessary, if not sufficient, condition for effectively protecting one's online privacy.
Individuals can protect their online privacy by adapting their behavior and through privacy-enhancing technologies (PETs).

\paragraph{Digital and Privacy Literacy}

\textit{Digital literacy} refers to the knowledge and technical skills to understand and use digital technologies and online content~\cite{Park2013,Peng2022}. Peng and Yu reviewed 20 studies that incorporate measurements of digital literacy, thereby revealing a multitude of approaches, each adapted to a specific purpose~\cite{Peng2022}. In particular, their survey highlights practices that include reusing existing scales, refining refining existing scales to specific contexts, and creating new ad hoc scales. 
In contrast, \textit{privacy literacy} refers to the knowledge about the collection and use of personal information~\cite{Park2013, Givens_2014}. Numerous instruments have also been developed in an attempt to measure this construct. Prominent among them is the Online Privacy Literacy Scale (OPLIS)~\cite{Trepte2015}. Such scales, however, can be onerous to apply and require significant contextual adaptations. For example, one of the six dimensions of OPLIS relates to the German regulatory framework for data protection. Büchi et al. reported on the relationship between digital literacy and privacy literacy, indicating that digital literacy is critical to understanding online privacy threats and protection measures~\cite{Buchi2017}. Much prior work has focused on assessing and improving privacy literacy, as there is evidence to show that both low privacy literacy~\cite{Liao2019} or overconfidence in one's privacy literacy~\cite{Jiaxuan2025} can result in people discounting privacy risks. 

\paragraph{Privacy-Preserving Online Behavior}

Although greater privacy awareness inspires privacy-preserving online actions, Kang et al. found that users tended to rely on their personal experiences instead of technical knowledge when making privacy-related decisions~\cite{KDF2015a}. One means to protect online privacy is to use discretion when browsing online, e.g. by being cautious of using social media and websites by public authorities for finding health-related information~\cite{Mitsutake2024}. A specific type of privacy-protection behavior is \textit{self-censorship}. Previous survey data suggests users may consider it as a protective measure~\cite{McDonald_Cranor_2010}, but it is influenced by nuanced dynamics with other factors, such as users' privacy concerns and perceived information value~\cite{Kim2025}. To develop a motivation to protect themselves, users must be aware that their privacy needs protection. In principle, users can become aware of tracking on a website through a website's privacy notice. However, users seldom read privacy notices~\cite{OO2020a}, or even mention them when discussing their privacy considerations~\cite{Jiaxuan2025, Kyi2024}. To motivate users to pay attention to their privacy, prior work has noted the potential of tools that clarify privacy details, such as by surfacing privacy risks~\cite{Shiri2024} and making privacy notices more concise~\cite{Ebert2021, Windl2022}. Users can also be encouraged to change their behavior by an increased awareness about the potential direct harm to them~\cite{Smith2024}. Going beyond basic protection, Masur proposed a vision of privacy literacy that also  ``motivates individuals to critically challenge the social structures and power relations that necessitate the need for protection in the first place''~\cite{Masur_2020}.

\paragraph{Privacy-Enhancing Technologies}

Privacy-enhancing technologies (PETs) are a class of software systems that can help users and administrators limit unwanted data collection or protect access to sensitive information~\cite{Seamons_2022}. Internet users can directly leverage privacy awareness and protection tools, such as Privacy Badger~\cite{PrivacyBadger} or NoTrace~\cite{MPS2013a}, when browsing the web. By receiving immediate feedback on the number of trackers on a website, users have the means to infer which websites track them and to what extent. 
However, many PETs are applications that users must actively seek, install, and activate. Because of the complexity of their underlying mechanism, people may have misconceptions about privacy-related web browsing tools, such as virtual private networks (VPNs) and private browsing~\cite{SSY2021a}. Some PETs can also mislead users into a false sense of security~\cite{Windl2025}. Common reasons for these misconceptions include users' partial knowledge~\cite{SSY2021a} and their resignation of the lack of privacy~\cite{Fong2023}. Prior work has focused on the use of such technologies for general web browsing. We contribute an understanding of how users leverage these technologies particularly in the context of seeking health information online.

\subsection{Research Questions}
\label{ss:ResearchQuestions}

We seek to better understand whether and how Internet users exercise three potential means of protection against implicit data collection when seeking health information online (RQ1--RQ3), and how they understand how this implicit data collection takes place (RQ4). 

\begin{description}
\item[RQ1] \textbf{How, if at all, do users consider the potential impact of the choice of website on their privacy when seeking online health resources?} As presented in Section~\ref{ss:HealthPortals}, different websites exhibit important differences in the amount of surveillance they deploy. We sought to understand to what degree users are aware of these differences and whether these differences matter to them.
\item[RQ2] \textbf{In which ways do users engage with privacy-enhancing technologies (PETs) to attempt to protect their privacy when seeking health information online?} A number of PETs are freely and widely available to help users protect their privacy. We sought to probe how users employ PETs and whether their use of PETs is purposeful and effective and results in a sense of protection.
\item[RQ3] \textbf{How do users consider the sensitivity of their inquiries and perform self-censorship when seeking health information online?} In our context, self-censorship corresponds to selectively limiting one's search for health information to avoid leaking personal information. 
We sought to understand whether and how users self-censor to protect their privacy in the context of health information. 
\item[RQ4] \textbf{How do users understand how their personal data is collected when they visit health information websites?} In particular, we sought to probe to what extent the third-party web tracking model described in Section~\ref{ss:WebTracking} was clear to health information seekers.
\end{description}

We developed research questions RQ1--RQ3 following a deductive strategy, by mapping each question to a different approach to privacy protection. In contrast, we developed RQ4 inductively during the data analysis.
\section{Study Method}

We conducted individual qualitative interviews with 35 adults who are residents of Canada. We restrict our scope to Canadian residents to be able to interpret the behavior and considerations of individuals seeking health information in a shared and well-understood context. In Canada, health care is provided through a universal and publicly-funded system that covers most healthcare needs. Canada's privacy legislative framework is built around two main federal laws: the Personal Information Protection and Electronic Documents Act (PIPEDA), which governs private-sector data handling, and the Privacy Act, which applies to federal government institutions. Some provinces have their own laws which can take precedence over federal laws. Compared to the GDPR, Canadian privacy legislation can be seen as offering weaker protection for individuals.

Our study was organized in two phases.
In a preliminary, \textit{exploratory phase}, we interviewed 15 participants to learn about their approach to seeking health information online and how they engaged in any data protection attempt, as per our first three research questions.
We leveraged the outcome of this phase to refine our data collection instruments and thereby enable a more focused inquiry in a second, \textit{directed phase}, with 20 participants.
In particular, we introduced an \textit{intervention} during the interview, consisting of a brief slide presentation in which we clarified important online surveillance concepts and mechanisms that had been overlooked or misunderstood by multiple participants in the exploratory phase. 
This protocol was reviewed and approved by the Research Ethics Board of McGill University.

\subsection{Participant Recruitment}

We recruited participants from departmental mailing lists at McGill University, as well as through social media, poster advertisements, and personal contacts. 
The study was open to all residents of Canada aged 18 years or older. Prospective participants were required to complete an online consent form and questionnaire to be invited for the interview. We recruited participants separately for each phase.

The pre-interview questionnaire required prospective participants to answer basic demographic questions (e.g., age, gender, educational achievement) using generalized categories, as well as a number of questions requiring them to self-assess their level of digital literacy, their attitude regarding privacy, and their knowledge of privacy threats. We refer to these last three factors as \textit{Background Attributes}. We used a different instrument to elicit background attributes in each phase of the study. 

For the exploratory phase, we derived our questions about \textit{attitude towards privacy} using an ordinal scale adapted from Elueze and Quan-Haase's typology of privacy attitudes~\cite{EQ2018a}. For digital literacy, we used a simple ordinal scale. We designed this scale to focus on participants' level of comfort in using software applications. The complete questionnaire is available in Appendix~\ref{a:s1Questionnaire} and the mapping between answers and scale levels is available in Appendix~\ref{a:mapping}. We received 20 completed questionnaires and recruited 15 of the 20 respondents to participate in an online video-interview.

The questionnaire responses we collected during the exploratory phase showed that our questions were only marginally discriminating. For example, all but one of the 15 participants indicated that they considered themselves ``Private in some way'' (the third level in the scale). Consequently, we revised our questionnaire to include additional questions, and relied instead on a subset of the questions from the 2022--2023 Survey of Canadians on Privacy-Related Issues, conducted by the Office of the Privacy Commissioner of Canada (OPC)~\cite{survey2023}. Our complete revised questionnaire is available in Appendix~\ref{a:s2Questionnaire}. 

For the directed phase, we wished to ensure a diversity in both demographic and background attributes. For this purpose, we used a combination of convenience and theoretical sampling and opted for a recruitment target of 20 participants. From our initial recruitment campaign, we received 107 completed questionnaires over several weeks. We initially recruited participants arbitrarily among the first respondents and proceeded with respondents with an unseen combination of gender, age group, education level, privacy attitude, and level of digital literary (as defined in Section~\ref{ss:Participants}). We recruited 13 participants in this way. However, all but two of these participants were respondents aged 18--34. We thus conducted another recruitment campaign targeting people 35 and older, and received an additional eight responses, for a total of 115. We invited seven of the eight respondents in this second cohort to achieve our target number of participants, for a total of 35 participants across both phases. The participants were compensated with a \$25 CAD gift card for completing the interview.

\subsection{Participants} 
\label{ss:Participants}

Table~\ref{t:Participants} provides the relevant self-reported characteristics of our 35 participants as collected through our pre-interview questionnaires. We summarize the data for each participant into a \textit{participant profile} that consists of two components: \textit{demographics} (Gender, Age, Education) and \textit{Background Attributes} (Privacy Attitude, Digital Literacy, and Threat Awareness).
For ease of interpreting a participant's background, we mapped all attributes to the uniform three-level scale: Low~\spie{0}, Moderate~\spie{180}, and High~\spie{360}. 
For the exploratory phase participants, we used the answers to some questionnaire questions directly, whereas for the participants of the directed phase, we used a system of rules to map from their questionnaire answers to the background attributes.
We provide the exact mapping procedure in Appendix~\ref{a:mapping}.

In consequence of our theoretical sampling, our set of participants exhibits a diversity of demographics and relevant backgrounds.
Specifically, we have participants from different genders, from all generalized age groups except one (45--54), and from all generalized educational levels. In terms of the summary background attributes, our set of participants covers 14 different configurations of values (out of 27 possible), and we verified that the attributes are not strongly correlated.\footnote{Pearson rank correlations, df=33: \textit{Privacy Attitude} vs. \textit{Digital Literacy} 0.17 (p=0.33); \textit{Privacy Attitude} vs. \textit{Threat Awareness} 0.12 (p=0.51); \textit{Digital Literacy} vs. \textit{Threat Awareness} 0.30 (p=0.079).}

The background attributes in Table~\ref{t:Participants} must be interpreted with care because segmenting users in terms of their privacy attitudes (e.g., as proposed by Westin~\cite{Westin2003}), is fraught with challenges. Among others, privacy concerns are contextual~\cite{Martin_Nissenbaum_2016} and prior research has shown that self-assessment of privacy literacy can be unreliable~\cite{Ma_Chen_2023}. In our own interview data, we also noted discrepancies between some participants' self-reported background attributes and our inference of these attributes. For transparency, we report the participants' self-reported values as a source of preliminary background, but rely on the interview data to interpret the participants' actual background.

\begin{table}[t!]
	\caption{Study Participants.}
	\label{t:Participants}
	\centering
	\begin{tabular}{lcccccc}
	\toprule
    & & &                                                             & \textbf{Privacy}  & \textbf{Digital}  & \textbf{Threat}\\
	\textbf{ID} & \textbf{Gender} & \textbf{Age} & \textbf{Education} & \textbf{Attitude} & \textbf{Literacy} & \textbf{Awareness}\\ \midrule
    \multicolumn{7}{c}{\textit{Exploratory Phase}} \\ \midrule
	\p{1}  & \female & 25--34 & Bachelor's  & \pie{180} & \pie{360} & \pie{180} \\
    \p{2}  & \female & 18--24 & Bachelor's  & \pie{180} & \pie{180} & \pie{180} \\
	\p{3}  & nb      & 25--34 & Bachelor's  & \pie{180} & \pie{360} & \pie{180} \\
	\p{4}  & \female & 25--34 & Bachelor's  & \pie{180} & \pie{360} & \pie{180} \\
	\p{5}  & \male   & 25--34 & Bachelor's  & \pie{180} & \pie{180} & \pie{180} \\
	\p{6}  & \female & 18--24 & College     & \pie{180} & \pie{0}   & \pie{180} \\
	\p{7}  & \male   & 18--24 & Master's    & \pie{180} & \pie{360} & \pie{360} \\
	\p{8}  & \female & 18--24 & High School & \pie{180} & \pie{180} & \pie{180} \\ 
	\p{9}  & \female & 25--34 & Master's    & \pie{180} & \pie{180} & \pie{180} \\
	\p{10} & \male   & 18--24 & Master's    & \pie{180} & \pie{360} & \pie{360} \\
	\p{11} & \male   & 25--34 & Bachelor's  & \pie{180} & \pie{360} & \pie{180} \\
	\p{12} & \male   & 25--34 & Master's    & \pie{180} & \pie{180} & \pie{360} \\
	\p{13} & \female & 18--24 & High School & \pie{180} & \pie{360} & \pie{180} \\
	\p{14} & \female & 25--34 & Bachelor's  & \pie{0} & \pie{360} & \pie{180} \\
	\p{15} & \female & 25--34 & Bachelor's  & \pie{180} & \pie{180}   & \pie{0} \\ \midrule
    \multicolumn{7}{c}{\textit{Directed Phase}} \\ \midrule
    \p{16} & \male & $\ge$65 & Ph.D. & \pie{0} & \pie{360} & \pie{180} \\
    \p{17} & \female & 18--24 & Bachelor's & \pie{180} & \pie{360} & \pie{0}\\
    \p{18} & \male & 25--34 & Bachelor's & \pie{180} & \pie{180} & \pie{0}\\
    \p{19} & \male & 25--34 & Master's & \pie{360} & \pie{360} & \pie{180} \\
    \p{20} & nb & 18--24 & College & \pie{360} & \pie{360} & \pie{0}\\
    \p{21} & \male & 25--34 & College & \pie{0} & \pie{360} & \pie{0} \\
    \p{22} & nb & 18--24 & College & \pie{360} & \pie{360} & \pie{0} \\
    \p{23} & \female & 18--24 & High school & \pie{180} & \pie{180} & \pie{0} \\
    \p{24} & \male & 25--34 & PhD & \pie{180} & \pie{180} & \pie{0} \\
    \p{25} & \female & 18--24 & Master' & \pie{0} & \pie{0} & \pie{0} \\
    \p{26} & \female & 18--24 & College & \pie{360} & \pie{360} & \pie{180} \\
    \p{27} & \female & 35--44 & PhD & \pie{180} & \pie{360} & \pie{180} \\
    \p{28} & \male & 25--34 & College & \pie{360} & \pie{360} & \pie{360} \\
    \p{29} & \female & 55--64 & Master's & \pie{0} & \pie{180} & \pie{0} \\
    \p{30} & \female & 35--44 & Bachelor's & \pie{0} & \pie{360} & \pie{180} \\
    \p{31} & \female & 55--64 & Bachelor's & \pie{0} & \pie{180} & \pie{0} \\
    \p{32} & \female & 35--44 & Master's & \pie{180} & \pie{180} & \pie{0} \\
    \p{33} & \female & 55--64 & Master's & \pie{180} & \pie{180} & \pie{360} \\
    \p{34} & \female & 35--44 & Master's & \pie{0} & \pie{180} & \pie{0} \\
    \p{35} & \male & $\ge$65 & Master's & \pie{180} & \pie{360} & \pie{360} \\
	\bottomrule
	\end{tabular}
	\end{table}

\subsection{Interview Protocol} 
\label{ss:Protocol}

We followed a semi-structured interview format designed to encourage participants to speak as freely as possible about their experience seeking health information online.
The same investigator conducted all 35 individual interviews through a cloud-based videoconferencing software. 
We audio- and video-recorded the interviews. All the interviews were conducted in English. The 15 exploratory phase interviews lasted on average 30 minutes ($SD=5.0$~minutes), and the 20 interviews in the directed phase lasted on average 42 minutes ($SD=8.2$~minutes). 

The \textbf{exploratory phase interviews} were organized in four parts. After an \textit{(1) administrative overview}, the interviewer asked the participant about \textit{(2) information sources} they go to for medical advice and how they decide between them. The interviewer followed the initial question with prompts as necessary, e.g., \textit{``Please tell me about your thinking process for selecting this resource''}.
The interviewer then asked the participant about the \textit{(3) factors that influence their interaction with health websites}. This part of the interview was designed to probe what participants think about when browsing health information websites and how they might protect their privacy doing so. In this part of the interview, if the participant did not themselves bring up privacy or surveillance topics, the interviewer explicitly asked, at a suitable juncture, \textit{``Have you thought about, and if so, what are your thoughts about your privacy as you search for online medical information?''}.
In an optional fourth part of the interview, the interviewer asked the participant to \textit{(4) walk through a specific health information seeking experience} they had recently. When necessary, the interviewer leveraged this part of the interview to elicit more details about the participant's experiences.

After the exploratory phase, we revised our interview protocol to elicit additional engagement on the topic of privacy. The interview protocol for the \textbf{directed phase} was also organized in four parts. After the same \textit{(1) administrative overview}, the interviewer asked the participant to imagine that they suspect they have shortness of breath and asked them to \textit{(2) walk through a health information experience} that mimics their last health information search session. As the participant sought health information by visiting websites, the interviewer asked them to verbalize their thoughts. Then, the interviewer conducted an \textit{(3) intervention} by giving a short, scripted slide presentation on web surveillance technology. This presentation consisted of four slides, summarized in Table~\ref{t:Intervention}. Finally, the interviewer elicited the participant's \textit{(4) reaction} by asking what information from the presentation was new to the participant, and how they felt about the information provided.

\begin{table}[t!]
	\caption{Summary of the Intervention (Directed Phase)}
	\label{t:Intervention}
	\centering
	\begin{tabularx}{\textwidth}{>{\raggedright\arraybackslash}p{0.3\textwidth} >{\raggedright\arraybackslash}p{0.65\textwidth}} \toprule
\textbf{Slide Title} & \textbf{Content (Excerpts)} \\
\midrule
1. Loading a web page connects to third parties. & When you access a web page, other companies can know about this. The page can contain various components, such as analytics libraries, social media buttons, and placeholders for ads. As a consequence, when the page loads, the browser will connect with the servers of providers for these components.\\
2. Third parties collect data even if you don't provide it. & As soon as you access a page, information about your visit is collected. This data collection happens without any explicit action from you. Here are four kinds of information that can be explicitly collected...\\
3. Web browsing data can identify you personally. & Companies match the data they collect automatically when you connect to a page to information they already have and use algorithms to identify you. The details can be very complex and technical, but the bottom line is that they are effective. What you see here is a typical visitor profile view for an analytics company...\\
4. Surveillance has real impact. & Here are four potential impacts of tracking: Risk of identity theft and scams, differential pricing, behavior manipulation, and algorithmic discrimination. \\ \bottomrule
\end{tabularx}
\end{table}

\subsection{Ethical Considerations} 

To avoid a potentially negative interview experience, we excluded participants with cyberchondria or related symptoms. Following the informed consent page, the first question of our preliminary recruitment questionnaire was \textit{``Do you get anxious when browsing medical information?''}. A prospective participant's positive answer ended the questionnaire by indicating the study was not suitable. Of the 150 prospective participants who answered our recruitment campaign, a total of 15 declined to continue by responding ``yes'' to the question. 

In addition to the usual anonymity and confidentiality conditions afforded to research participants, we implemented a number of measures to ensure that their participation in the study would not result in a privacy invasion. In particular, we set up the videoconferencing system to allow the participants to browse the web on the interviewer's computer. This setup enabled a participant to explore websites and demonstrate web browsing actions without any tracking or other consequence to the state of their personal computer (such as additional cookies or entries in their browsing history). Additionally, during the administrative phase of the interview, we mentioned to the participants that they did not need to share any detail of their medical history as part of the interview. We instructed them to use ``asthma'' or ``hypertension'' as a placeholder if they wanted to discuss a prior personal experience. All 35 participants completed the interview, with none withdrawing during or after.

\subsection{Data Analysis}

We analyzed the interview transcripts using a hybrid deductive/inductive approach. In our qualitative analysis, we followed a systematic process of \textit{reduction} and \textit{classification}~\cite{Rou2014a}. This decision stems from our intention to initiate our inquiry by examining our first three well-defined research questions.

A single investigator completed a pass of open coding. Proceeding incrementally by research question, this investigator first identified all excerpts relevant to a given question. Then, the investigator qualitatively coded each excerpt using a small set of codes elaborated inductively, and tabulated the codes across all participants. The outcome of the coding phase was thus a matrix capturing, for all participants, a number of codes reflecting the participants' considerations relevant to a question, joined with the participants' profile as described in Section~\ref{ss:Participants}.

As part of the analysis, we identified a new important research question, namely \textit{RQ4: How do users understand how their personal data is collected when they visit health information websites?} We then re-analyzed the transcripts in the same way as for our three a priori questions. We leveraged the four \{$\mathrm{code} \times \mathrm{participant}$\} matrices thus produced through our qualitative coding as the basis for interpreting the participants' approach and considerations. Although the qualitative coding was done on a per-question basis, it relied on an interpretation of the complete interview with each participant, including taking into account the participant's profile. As the code matrices the investigator created were a part of the analysis process and neither a measurement nor the outcome of the analysis, we judged the application of inter-rater reliability assessments to be unsuitable~\cite{MSF2019a}. Instead, we sought to improve the reliability and credibility of the analysis through an iterative process of review and discussion among three experienced investigators. 

\section{Results}
\label{s:privacyConsiderations}

This section reports on answers to our four research questions. We first synthesize how our participants considered, or not, three different means of control to protect their privacy while seeking health information online (RQ1--RQ3). We follow up with an analysis of how our participants thought about the surveillance of their browsing activities, in particular by third parties (RQ4).

\subsection{How Health Information Seekers Select Websites}
\label{ss:theme1}

All 35 participants described searching for information via a search engine, and then browsing the results.
This strategy is best summarized by \p{11} (\male~25--34 high digital literary): \q{to find anything related to health that I'm looking for online, [I] just Google some stuff}. An alternative to this strategy is \textit{direct access} of a usual or preferred website through its URL or a bookmark. However, none of our participants reported accessing heath websites directly. We thus investigated how participants select the results of their web searches on health topics. 

When we asked them about how they choose between search results, \textit{none of the participants indicated that they select websites based on privacy protection goals}. Multiple participants expressed either selecting among the first search results or general ambivalence about the source of the health information they sought. As \p{2} (\female~18--24 moderate digital literacy), told us: \q{Usually I will just look up whatever I'm looking for [...] I won't necessarily look for a website and then go on the website that I know is trusted...}. For our participants, meeting their informational need trumped loyalty to any given website. In fact, some participants, including \p{2}, even indicated that they had never thought about the source of their information before participating in the study. 

Instead, most participants mentioned \textit{familiarity} with a website and an impression of \textit{authoritativeness} as the main criteria for selecting a website. In the words of our participants, \textit{familiarity} referred to whether they had seen a website before, whether they ``recognized'' it, or whether they believed it was ``popular'': \q{most of my common sources include WebMD because that's the most popular}\p{12} (\male~24--34); \q{Cleveland Clinic, I recognize that, I think I can trust that. MayoClinic, I recognize that as well.} \p{22} (nb~18--24), etc. A number of participants mentioned selecting websites specifically because they thought they seemed ``reliable'', ``legitimate'', or ``trustworthy'', which we jointly refer to as \textit{authoritative}. We found that our participants' notion of authoritativeness could be superficial and based on simple clues such as the presence of ``MD'' (medical doctor) in the site's branding, or a subjective personal opinion. This informal process for assessing websites was best characterized by \p{10} (\male~18--24, high digital literary): \q{there are some websites that are more professional than others and that seem more trustworthy, but it's hard to tell sometimes}. 
When participants talked about \textit{trust}, they were referring to their belief that the information provided by a website was accurate, as opposed to a concept of trust related to privacy protection~\cite{Knowles2023}. Finally, we observed that the concepts of familiarity and authoritativeness can easily be conflated: participants assume that websites are authoritative because they know them. As \p{20} (nb~18--24, highly private) told us: \q{those [websites] I kind of trust, but I would have no proof or reasoning or anything to tell you why they're trustworthy. [...] It's just: these are websites that I've used before, I'll go for it}. 

Participants discussed \textit{other strategies for choosing health resources}, but to a lesser extent than familiarity and authoritativeness. A few participants mentioned paying attention to some of the attributes of the content, such as how accessible the text is (\p{13}, \p{18}) or avoiding a particular website because the content is \qq{too dramatic}{20}, or not wanting to see \qq{anything medical}{22} (in the sense of technical), etc. Nine participants also mentioned looking specifically for experiential health information (typically on the forum Reddit), a practice outside the scope of the present study. A few participants mentioned using a generated summary or a conversational agent. Finally, seven participants  mentioned opening multiple websites to cross-check information. Participant \p{14} (\female~25--34 high digital literary), best described this practice: \q{I look and then I just cross check with other sites just to see if it's congruent or not}.
In the context of a study on the collection of web browsing information, this practice is noteworthy. First, accessing multiple websites risks disseminating a user's browsing information to an even greater number of data collectors. Second, some health information on the web is syndicated, meaning that multiple websites may show content from the same source. For instance, both Healthline, the Center for Disease Control, and the National Institute of Health offer syndication of their content. 

\paragraph{Summary:} Participants did not discriminate between health websites based on privacy protection goals. Instead, they prioritized the nature of the content and a subjective impression of the website's authoritativeness.

\subsection{How Health Information Seekers Use Privacy-Enhancing Technologies}
\label{ss:theme2}

Users can adjust settings and use features of their web browser or additional tools to protect their privacy. Such mechanisms are jointly referred to as \textit{privacy-enhancing technologies} (PETs). About half our participants discussed either using PETs or being aware that PETs were deployed on their system when they browsed the web.

Seven participants mentioned they \textit{rejected cookies}, although not in relation to health websites or specific concerns about privacy. Rather, most participants who mentioned rejecting cookies did it as a matter of routine. Only one participant, \p{30} (\female~35--44 non-private, high digital literacy) discussed some form of discrimination when denying cookies: \q{I barely have paid attention to them unless the website which I trust, I accept cookies, but there are many websites I don't accept the cookies. But I don't have much knowledge on them}. This participant's comment illustrates the fact that that pop-up cookie banners force users to make a decision regarding cookies whether they understand them or not~\cite{SSY2021a}. Additionally, a user's decision may be significantly impacted by the design of the cookie banner~\cite{BLN2024a}. Multiple participants shared a sense of cynicism about denying cookies. This sentiment was best exemplified by \p{15} (\female~25--34 low privacy threat awareness): \q{I just deny, but I don't know how much that protects me from being tracked.}

Similar to denying cookies, some participants mentioned using a browser's \textit{private mode}, but again without a conviction of its effectiveness at protecting their privacy.
For example, \p{5} (\male~25--34), described his attempt at data protection as \q{I try to go Incognito sometimes, delete browser cookies or things like that, but... [it's] like a shallow thing to do.}. This sentiment was echoed by \p{9} (\female~25--34): \q{I wouldn't be surprised if I've gone on private mode or something, but I don't even know: would that, like, help?}

Seven participants discussed the use of \textit{ad and tracker blockers}. However, in discussing the use of ad/tracker blockers, our participants' verbalizations indicated a passive use of the technology, of something running in the background, similar to antivirus software: 
\qq{and so I started using the adblocker that comes with the browser Firefox}{26}. Only two participants mentioned installing external tracker blockers themselves (respectively uBlock Origin (\p{22}) and Ghostery (\p{24})), and none of our participants mentioned adjusting settings in tracker blockers, and only one mentioned looking at the easily accessible reports from such tools.

\paragraph{Summary:} While some participants used PETs, this usage lacked a conviction that the technology made a difference for them. None of the participants discussed making use of PETs to protect information about their medical history or interests.

\subsection{How Health Information Seekers Self-Censor}
\label{ss:theme3}

We investigated whether and how the participants would selectively limit their search for health information to avoid leaking personal information.

When discussing their general process for seeking health information online, most of our participants did not describe \textit{any concrete experience with self-censorship}. In particular, their description of the process they use to find relevant information, including the selection of query keywords, included no provision for adjusting or omitting some terms. To stimulate the participants' memory and avoid oversights, the interviewer systematically used a prompt for broaching the topic. In the exploratory phase, this prompt was \textit{``are there any differences in how you interact with websites when it comes to health information and non-health information?''}. This prompt was designed to get participants to think about whether and how they might search differently given the sensitivity of the topic. Fifteen participants did not consider the possibility of self-censorship in response to this question, sometimes not understanding the question, or referring to the use of privacy-enhancing technologies instead (\p{13}, \p{28}). For example, \p{20} (nb~18--24 private, high digital literacy), engaged with this question by discussing how they mentally categorize their need for information: \q{I'm thinking of three broad categories. If I have something that I'm really worried about, if I have something that I think might be an emergency or something's just really wrong, I will look it up to see what it could be, how urgent it is...}. An additional 13 participants understood the reference to self-censorship and simply answered that they did not engage in this practice: \qq{I search everything the same way}{21}. For some of these latter participants, the reason they did not self-censor was a combination of general privacy attitude and the apparent absence of threats. For example, \p{2} and \p{8} (both \female~18--24 moderately private), told us: \qq{I would say I honestly feel like my medical information: I'm fine with sharing that information, I feel comfortable with it}{8};  \qq{I don't think I have any big conditions that I could [...] suffer from sending out in the world}{2}.

Six participants engaged with our question on searching differently, but these participants \textit{considered self-censorship in a speculative way} by discussing hypothetical scenarios as opposed to reflecting on their experience. These six participants all used the subjunctive mood to describe their views: \qq{I wouldn't put it on the Internet}{7}; \qq{if it was something more sensitive, I may not look it up at all}{19}. Understandably, their justification for an eventual self-censorship centered around avoiding unspecific harm, e.g., \qq{Whereas [leaking] health data could have a big effect on you}{10}.

Ultimately, only a single participant described a personal self-censorship behavior in concrete terms. Participant \p{31} (\female~55-64 non-private) mentioned \q{...if I'm doing my own search and it's about suspected medical condition that I'm embarrassed about, I might not even do the search, but if it's a condition that is a common medical condition, then I tend to not filter my searches}. 

\textbf{Summary}: Self-censorship is not a practice our participants used. This data protection technique was not actively employed by, or had not even occurred to, all but one participant.

\subsection{How Health Information Seekers Understand Web Tracking}
\label{ss:theme4}

While most participants were aware that their data \textit{was collected}, few could articulate clearly \textit{how} and \textit{by whom}. Many people have limited knowledge of how the Internet and the surveillance economy work and, not suprisingly, this limited knowledge can affect how they make privacy-related decisions when browsing the web~\cite{KDF2015a}. We examined how our participants construed implicit data collection, in particular by third-party organizations.

Most of our participants referred to the collection of their personal data exclusively in \textit{abstract terms}. In discussing the data collection process, their verbalizations supressed the agent or used generic pronouns such as ``they'', suggesting a lack of clarity about who exactly does the data collection. For instance, \p{6} (\female~18--24) referred to \q{the information that would be saved about me on the Internet}. Such generic references to web tracking extended to the participants description of their use of PETs: \q{I know that anytime you're online, even if you're on Incognito, it's [recorded].} \p{23} (\female~18--24). As we heard from \p{24} (\male~25--34~PhD), users can be aware of web tracking yet ignore the process: \q{[I] feel that everything I'm doing online is always [going to be] tracked and logged in some way. [...] But there seems to be a baseline of anytime you get on the Internet, something is being extracted about my behavior, but I don't know what it is}. When participants mentioned specific data collectors, these consisted mostly of the popular platforms they directly interact with: Facebook/Meta, Google, Tiktok. When participants expressed awareness of third-party data recipients, their awareness of tracking was linked to cookie banners and targeted ads, namely two obvious artifacts of web tracking activities that can unsettle web users~\cite{RWM2024a, ULP2012a}. 
Only four participants discussed the collection of their data by data brokers or advertisement companies: in detail by three participants with high digital literacy (\p{19}, \p{21}, \p{26}), and by \p{6}, who mused about how her data was used: \q{I think what I would want to know is if they sell [my information] and who are they sending it to? And those groups that they are sending to, what are they doing with it? Are they  advertisement companies?}.

About one third of our participants referred to their data being viewed or accessed by \textit{individual persons}. Although a threat does exist that an online health seeker's browsing history could be abused by unauthorized persons or for inappropriate purposes such as harassment~\cite{Jackson2021}, most web browsing data is processed algorithmically. In this context, reference to individuals as data collectors could be evidence of a divergence between a participant's mental model of web tracking and the reality of the surveillance economy. In a number of cases, references to persons could be interpreted as an abstraction of the data collector, as we discussed above, e.g., \qq{what if if someone had access to my information, what are they going to do with it?}{1}. However, in a number of instances, it was clear that the participant was thinking of violations of their privacy specifically by an individual: \q{I take for granted that anything I look at could be accessed by someone with sufficient determination to get there} \p{26}(\female~18--24).

When participants referred to data collectors specifically instead of in abstract terms, they mainly \textit{focused on the first party}: they talked about surveillance from ``the website'' as the entity they interacted with directly. 
For these participants, it is ``the website'' that is doing the tracking: \q{It's not like, if I look for symptoms of asthma online and the websites are going to track me more compared to other website} \p{12}(\male~25--34). As another example, \p{11} considered the first party as the only data collector and might have overlooked implicit data collection: \q{[If] the website [is] from a source that I don't even trust at all, then I'm not leaving any information at all} \p{11}(\male~25--34), an issue reflected almost identically by \p{14} (\female~25--34): \q{[Privacy]'s never been a concern because the websites I go to never ask for personal information}. Even the better-informed participants, who distinguished between the concepts of a website and its operator, exhibited a perspective whereby the flow of their data is mediated by the first party exclusively: \q{Look at the website, the agents behind it, their affiliation, look at what they might do with your data.} \p{10}(\male~18--24, high digital literacy).

As part of our analysis of how participants understood web tracking, we leveraged the intervention we used in the directed phase to take into account the participants' reaction to the intervention, in which we presented the third-party data collection model present on most websites (see Section~\ref{ss:Protocol}). 

In response to our intervention, eight of the 20 participants told us that \textit{they had not known about third-party data collection.} This list of participants includes two with a high self-reported level of digital literacy, e.g., \q{New to me was, [...] why would third-party websites be loaded within a [first party] website?} \p{19}(\male~25--35). An additional participant, \p{21}(\male~25--34), insisted that he was aware of how third parties collect data from web browsing, but elaborated in a way that contradicted his statement: \q{People are always trying to escape from [tracking] by using open source applications}. What we heard from the remaining 12 participants was not a clear indication that they were fully cognizant of the third-party data collection model. Although these latter participants did not indicate that they had learned anything new from our slide on third-party tracking, their discussion with the interviewer exhibited many of the types of behavior we described in this section, and in particular focusing exclusively on the first party and using abstract terms such as the generic ``they'' to refer to data recipients. For instance, when asked \textit{``Were there a type, or types, of information that you didn't know was collected?''}, \p{24} (\male~25--34~PhD) replied \q{I'm not that surprised by what they collect. It's just [that], before I had the understanding that when I went on to any given site, it was kind of starting at zero [...] So when I go in healthline.com or something, if it's the first time I've ever been to Healthline, my understanding is that I'm starting at zero with Healthline. And so they only get from me whatever I give them.} This verbalization is just one of many examples of how the ``first party'' model of data collection persists in the mind of health information seekers. For many of our participants, their reaction to our presentation on third-party collection seemed intuitive. However, despite being aware of tracking, multiple participants exhibited an inaccurate mental model of how this tracking occurs in practice. These inaccuracies included surveillance via the browser, a ``bad actor'' model of surveillance, or tracking scenarios limited to search engines or via authentication through services. 

\textbf{Summary}: When seeking health information online, our participants either were not aware of data collection by third parties or could not articulate precisely how implicit data collection by third parties takes place.

\section{Discussion}
\label{s:discussion}

We contextualize the findings, suggest new directions for the support of surveillance awareness and privacy education, and reflect on the limitations of the study.

\subsection{Theoretical Contextualization}

The overarching observation we can draw from our study is that participants did not indicate investing much, if any, effort in protecting their privacy when seeking health information online (RQs 1--3). In fact, many participants seemed not to have considered the implicit collection of their personal information by third parties as part of this process. The dominant factor in their search was the quality and relevance of the information. At the same time, all participants were aware that information about them is collected when they access websites. 

In the logic of the privacy calculus, our interpretation is that participants ascribe a higher value to the information they obtain from health websites than the perceived cost. Indeed in the context of health information seeking, a user anxious about their health, or the health of a close one, is highly motivated to obtain relevant information. We surmise that urgency, combined with the value we ascribe to our health, leads  
to a perception of high relative value for the information on health websites. In contrast, the practical implications of privacy violations are diffuse and intangible. In Canada, where a universal healthcare system is in place, major implications of surveillance such as price discrimination or insurance denial do not apply directly and may contribute to a decrease in the perception of risk, and thus of the relative ``cost'' of privacy invasions. A potential confounding factor is that most information collection on health websites is implicit. With the routinization of  personalized ads and cookie banners, implicit data collection becomes background noise, with the potential consequence that if participants do not consider their web search activities as information disclosure, there is no apparent ``cost'' to their use of health websites. This perspective could potentially explain why participants did not consider the possibility that some websites track them less than others as a way to exercise control over their privacy (RQ1).

From our investigation of how users understand the collection of their personal data on information websites (RQ4), we conclude that for most participants, both the data collection mechanisms and data flow paths are opaque. In particular, without a purposeful use of tracker blockers (RQ2), users may simply not be aware of the extent to which they are tracked. According to the protection motivation theory, users would be motivated to protect their privacy when visiting health information websites if they could properly assess the threat and determine how to cope. We surmise that both \textit{threat appraisal} and \textit{coping appraisal} are degraded in the face of the obscure mechanisms whereby visitor data flows to third parties. In that sense, our findings complement previous observation by Boerman et al.~\cite{Boerman2021} that users are not motivated to protect their online privacy. Although the threat appraisal can be difficult, the assumptions motivating this research is that coping is not technically difficult. In particular, different health information websites that could offer comparable information exhibit large differences in the degree to which they track users. However, our participants' health information seeking behavior indicates that they ignore such differences (RQ1).

In terms of the contextual integrity lens, we derived our most useful insights from our investigation of whether participants self-censored (RQ3). Among our participants, this was not the case. The fact that they \q{search everything the same way} (Section~\ref{ss:theme3}) may indicate the presence of an established norm for web search. When seeking health information, participants may simply not feel like an important context switch takes places between health-related and more mundane web searches. While users consciously adapt their behavior when explicitly sharing information online (see Section~\ref{ss:SeekingHealthInformation}), they did not appear to do so when accessing websites that only collect data implicitly.

\subsection{Practical Implications}

While our findings highlight the obstacles to privacy protection for health information, they also point out directions to help combat unwanted data collection by third parties. To increase the data self-protection and informational self-determination, health information seekers, and Internet users in general, need to be informed of how data exactly flows to whom. 

As Nissenbaum observed, the proliferation of systems that collect personal information occurred jointly with ``diversification in the social actors who maintain and use them''~\cite{Nissenbaum2010}. Ignoring the third-party collectors would overlook this diversity of aims and motivations. For an Internet user who considers that their personal data \q{is taken} by \q{the website}, it is understandably difficult to imagine the distinction between a first party that only conducts their own data collection for analytics purposes, from a government website that sends data to Google to leverage the convenience of their analytics platform, from a website designed to optimize the monetary value of visitor traffic. Hence, in the terms of the contextual integrity theory~\cite{Nissenbaum2010}, web users cannot determine whether implicit data collection is appropriate and justifiable. For example, after the intervention, \p{19} asked \q{why would third party websites be loaded within a WHO website?}. To properly answer this question, Internet users need to grasp some of the political and economic forces that influence website operators. This calls for more transparency from the website to help Internet users to establish trust~\cite{Tsirozidis2025} that is not just ``hopeful''~\cite{Knowles2023}.

Lack of awareness of how third-party data collection works can also hamper users' confidence in their own agency. Our analysis includes rich descriptions to explain the phenomena. By describing data collection with the passive voice and data collectors with the generic ``they'', participants expressed their lack of agency and control. Many participants explicitly stated, after the intervention, that they did not plan to change anything about their behavior because of their impression that they could not do anything about data collection. In reality, some first parties can probably be trusted, and therefore one simple means of control over one's data is to access websites that at least do not immediately leak connection information, such as the UK's NHS's (see Table~\ref{t:PortalContent}). To allow users to distinguish between websites and their numerous competitors, we need more standardized and accessible methods to compare the websites' tracking practices, without placing the burden on users to find and meticulously review privacy notices. 
Our study further suggests a need for a foundational educational curriculum that includes how personal information is harvested by different information systems.

\subsection{Limitations}

Our study exhibits two limitations that are related to the generalizability of the findings. First, we targeted our recruitment efforts at engaging with participants with different personal characteristics. Despite these efforts, most of the volunteers we recruited have a university degree, which provided us with relatively fewer perspectives from Internet users with a different educational profile. For this reason, we do not expect that the considerations we report will exhaustively cover the set of privacy-related considerations of online health information seekers. However, in many respects, our participants exhibited homogeneous behavior and attitudes with respect to seeking health information online, despite their personal differences in background. By focusing on perspectives, attitudes, and behavior shared or corroborated by multiple participants, we ensure that the considerations we report on are not idiosyncratic.

A second noteworthy limitation concerns our mid-study adjustment of the method, including an updated instrument to characterize the participants' background and a slightly different interview guide. Although this methodological decision was intended to increase the usefulness of the data we collected, it involves a trade-off that must be accounted for when interpreting the results. The use of different instruments for assessing the participants' background attributes means that these are not directly comparable across phases. For this reason, we have avoided their use for any analytical purpose except a basic control for correlation, instead relying on them only as a starting point for understanding our participants' background and better interpreting the interview data. Although our two interview guides addressed the same topics, the use of questions and prompts phrased differently means that our participants engaged with our areas of inquiry in different ways. We remained sensitive to this difference throughout the analysis process, and conducted the analysis by relying on a comprehensive review of each transcript to ensure quotes would not be taken out of context. As the goal of our interviews was to engage our participants in a conversation (as opposed to eliciting answers to precise questions), we determined that performing an integrated analysis on the interviews of both phases did not pose undue risks of misunderstanding or misreporting the data.
\section{Conclusion}
\label{s:conclusion}

Health information is available online from several websites with a similar structure and comparable content. Yet, our analysis shows that these websites exhibit large differences in the number of third parties that track visitors to record their browsing activities. Through interviews with 35 participants with different backgrounds and privacy attitudes, we observed limited initiative towards data protection when seeking health information online. Our participants did not discriminate between health websites based on privacy protection goals, their use of privacy-enhancing technologies sometimes lacked a conviction that it made a difference for them, and they did not self-censor for data protection. 
While most of our participants were conscious of data collection, many were not aware of implicit data collection by third parties or could not articulate accurately how this data collection by third parties takes place. Our observations suggest that one key to help users shield their browsing activities from surveillance is to become better informed about how their data is collected and by whom so that they can effectively discriminate between different information providers on the basis of how well they respect their privacy. Two ways to effect this improvement in awareness include more standardized and accessible methods to compare the websites' tracking practices and an education curriculum that includes how personal information is harvested by different information systems. Our hope is that if a majority of users are empowered to effectively discriminate between online services on the grounds of their privacy practices, web traffic could flow to alternatives that can meet this demand.

\subsection*{Acknowledgments}

The authors are grateful to their colleagues and to the anonymous reviewers for feedback on the study design and earlier versions of this manuscript. This work was supported by the Social Sciences and Humanities Research Council of Canada under Grant 430-2022-00182.

\subsection*{Declaration of Interest Statement}

The authors report there are no competing interests to declare. 

\bibliographystyle{plain}
\bibliography{references}

\appendix

\section{Health Information Websites Search Results}
\label{a:websites}

The following is the list of the top 50 heath websites retrieved using SimilarWeb\footnote{\url{similarweb.com}, accessed 2023-05-18.} with the filters \mc{health} and \mc{worldwide} on 18 May 2023, ordered by popularity (across then down). We then eliminated from the list the sites that either did not include English as a main language, that did not provide information organized by health topic, that focused on drug information rather than health topics (e.g. Drugs.com), or whose information was targeted to professionals (e.g., Medscape). 
 
\begin{center}
    \small
    \centering
    \begin{tabular}{>{\raggedright\arraybackslash}p{0.20\linewidth}>{\raggedright\arraybackslash}p{0.20\linewidth}>{\raggedright\arraybackslash}p{0.20\linewidth}>{\raggedright\arraybackslash}p{0.20\linewidth}}
    \toprule
    nih.gov & healthline.com & mayoclinic.org & webmd.com \\
    medicalnewstoday.com & cvs.com & clevelandclinic.org & cdc.gov \\
    medlineplus.gov & nhs.uk & msdmanuals.com & walgreens.com \\
    sportkp.ru & altibbi.com & 1mg.com & doctolib.fr \\
    menshealth.com & alodokter.com & tuasaude.com & aarp.org \\
    womenshealthmag.com & vinmec.com & halodoc.com & medonet.pl \\
    doctoralia.com.br & drugs.com & vidal.ru & activebeat.com \\
    verywellhealth.com & athenahealth.com & abczdrowie.pl & my-personaltrainer.it \\
    psychologytoday.com & who.int & apteka.ru & goodrx.com \\
    webteb.com & mscoldness.com & babycenter.com & fitbit.com \\
    health.clevelandclinic.org & hellosehat.com & vnimanie.pro & uworld.com \\
    everydayhealth.com & medscape.com & myfitnesspal.com & hopkinsmedicine.org \\
    eatthis.com & rlsnet.ru \\
    \bottomrule
    \end{tabular}
\end{center}
\section{Health Information Websites URLs}
\label{a:urls}

The following are the URLs of the websites in our dataset. For each website, we identified the directory page or pages for the articles available and aggregated the URLs of all articles into an index.
For a given index, we inspected the links to determine whether each article available through the index was related to a \textit{health concern}, defined as articles on a symptom, condition, or treatment related to a person's health. Although most websites provide a majority of articles on topics that conform to this definition, others include information on healthy lifestyle topics, such as ``Active Children''. We focus on health concerns due to their potentially sensitive nature. For each website, we identified any pattern in the link URLs that indicated a page falling outside of our definition and excluded URLs matching these patterns in the automatic extraction process. We then manually inspected the index and removed any remaining page whose content did not match our above definition of a \textit{health concerns}. 

    \begin{tabular}{lll}
    \toprule
    \textbf{Website} & \textbf{Type} & \textbf{URL} \\
    \midrule
    Canada & Canadian & \href{https://canada.ca/en/public-health/services/diseases.html}{canada.ca/en/public-health/services/diseases.html} \\
    Alberta & Canadian & \href{https://myhealth.alberta.ca/health/Pages/default.aspx}{myhealth.alberta.ca/health/Pages/default.aspx} \\
    British Columbia & Canadian & \href{https://www.healthlinkbc.ca/illnesses-conditions}{healthlinkbc.ca/illnesses-conditions} \\
    Manitoba & Canadian & \href{https://www.gov.mb.ca/health/publichealth/atoz\_diseases.html}{gov.mb.ca/health/publichealth/atoz\_diseases.html} \\
    Ontario & Canadian & \href{https://health811.ontario.ca/static/guest/medical-library}{health811.ontario.ca/static/guest/medical-library} \\
    Qu\'{e}bec & Canadian & \href{https://www.quebec.ca/en/health/health-issues/a-z}{quebec.ca/en/health/health-issues/a-z} \\
    CDC & Other Gov. & \href{https://www.cdc.gov/health-topics.html}{cdc.gov/health-topics.html} \\
    MedlinePlus & Other Gov. & \href{https://www.medlineplus.gov/all\_healthtopics.html}{medlineplus.gov/all\_healthtopics.html} \\
    NHS & Other Gov. & \href{https://www.nhs.uk/conditions/}{nhs.uk/conditions/} \\
    WHO & Other Gov. & \href{https://www.who.int/health-topics/}{who.int/health-topics/} \\
    Cleveland Clinic & Commercial & \href{https://my.clevelandclinic.org/health/diseases}{my.clevelandclinic.org/health/diseases} \\
    Everyday Health & Commercial & \href{https://www.everydayhealth.com/conditions/}{everydayhealth.com/conditions/} \\
    Healthline & Commercial & \href{https://www.healthline.com/directory/topics}{healthline.com/directory/topics} \\
    Mayo Clinic & Commercial &  \href{https://www.mayoclinic.org/diseases-conditions/}{mayoclinic.org/diseases-conditions/}\\
    Verywell Health & Commercial & \href{https://www.verywellhealth.com/health-a-z-4014770}{verywellhealth.com/health-a-z-4014770}  \\
    WebMD & Commercial & \href{https://www.webmd.com/a-to-z-guides/health-topics}{webmd.com/a-to-z-guides/health-topics} \\
    \bottomrule
    \end{tabular}
\section{Pre-Interview Questionnaire: Exploratory Phase}
\label{a:s1Questionnaire}

\begin{enumerate}
\item Do you get anxious when browsing medical information? \textit{Participation is not recommended if you are anxious about your health.}
	\begin{enumerate}
		\item Yes
		\item No
	\end{enumerate}

\item Full name \textit{last, first}
\item E-mail address \textit{The address that was used to contact us}
\item Verification code \textit{The code you received by email after contacting the investigators}

\item What is your age group? \textit{You must be 18 or over to participate.}
	\begin{enumerate}
		\item 18--24
		\item 25--34
		\item 35--44
		\item 45--54
		\item 55--64
		\item 65 or above
	\end{enumerate}

\item How do you identify in terms of gender identity? \textit{If you prefer to self-describe please use the last option to enter your preferred term}
	\begin{enumerate}
		\item Woman
		\item Man
		\item Non-binary
		\item Prefer not to say
		\item {[Other]}
	\end{enumerate}

\item What is your highest level of educational attainment?
	\begin{enumerate}
		\item High school
		\item Post-secondary college or equivalent
		\item Bachelor's degree or equivalent
		\item Master's degree or equivalent
		\item PhD or equivalent
	\end{enumerate}

\item What is your general attitude towards privacy?
	\begin{enumerate}
		\item I consider myself very private
		\item I consider myself private in some ways
		\item I do not consider myself private
		\item I don't know or cannot answer
	\end{enumerate}

\item What is your understanding of the risks of disclosing personal information when using digital media?
	\begin{enumerate}
		\item I am not aware of the risks
		\item I have a basic familiarity with the risks
		\item I am well informed of the risks
		\item I don't know or cannot answer
	\end{enumerate}

\item What statement best describes your attitude towards disclosing information when using digital media?
	\begin{enumerate}
		\item I am very reticent about disclosing information in most situations
		\item I don't like having to disclose information, but I sometimes need to make a trade-off between protecting privacy and engaging with digital media
		\item I don't mind divulging information about myself when using digital media
		\item I don't know or cannot answer
	\end{enumerate}

\item What is your highest level of digital literacy?
	\begin{enumerate}
		\item I can use simple software tools: a web browser, an email program, etc.
		\item I can use various software tools and internet services to do a range of tasks, but I am not an advanced user of any of them.
		\item I would consider myself an advanced user of one or more software technologies, but this is not my primary field of activity.
		\item I am an IT professional or student.
	\end{enumerate}

\item Why do you want to participate in this study?

\item Please provide us with a Canadian mailing address. \textit{We need this information to validate that you are a resident of Canada and to send you the gift card.}
\end{enumerate}
\section{Pre-Interview Questionnaire: Directed Phase}
\label{a:s2Questionnaire}

\begin{enumerate}
\item Do you get anxious when browsing medical information? \textit{Participation is not recommended if you are anxious about your health.}
	\begin{enumerate}
		\item Yes
		\item No
	\end{enumerate}

\item Full name \textit{last, first}
\item E-mail address \textit{The address that was used to contact us}
\item Verification code \textit{The code you received by email after contacting the investigators}

\item What is your age group? \textit{You must be 18 or over to participate.}
	\begin{enumerate}
		\item 18--24
		\item 25--34
		\item 35--44
		\item 45--54
		\item 55--64
		\item 65 or above
	\end{enumerate}

\item What is your gender identity? \textit{If you prefer to self-describe please use the last option to enter your preferred term}
	\begin{enumerate}
		\item Woman
		\item Man
		\item Non-binary
		\item Prefer not to say
		\item {[Other]}
	\end{enumerate}

\item In which province or territory do you live?
\begin{enumerate}
    \item Alberta
\item British Columbia
\item Manitoba
\item New Brunswick
\item Newfoundland and Labrador
\item Northwest Territories
\item Nova Scotia
\item Nunavut
\item Ontario
\item Prince Edward Island
\item Quebec
\item Saskatchewan
\item Yukon
\end{enumerate}
 
\item What is your highest level of educational attainment?
	\begin{enumerate}
		\item High school
		\item Post-secondary college or equivalent
		\item Bachelor's degree or equivalent
		\item Master's degree or equivalent
		\item PhD or equivalent
	\end{enumerate}

\item What is your highest level of digital literacy?
	\begin{enumerate}
		\item I can use simple software tools: a web browser, an email program, etc.
		\item I can use various software tools and internet services to do a range of tasks, but I am not an advanced user of any of them.
		\item I would consider myself an advanced user of one or more software technologies, but this is not my primary field of activity.
		\item I am an IT professional or student in a field related to Computer Science.
	\end{enumerate}

\item How closely, if at all, do you follow news about privacy issues?
\begin{enumerate}
\item Very closely
\item Somewhat closely
\item Not too closely
\item Not at all
\end{enumerate}

\item In general, are you concerned about the protection of your privacy?
\begin{enumerate}
\item Extremely concerned
\item Concerned
\item Somewhat concerned
\item Not concerned
\end{enumerate}

\item How much do you agree with the following statement - "I am confident that I have enough information to know how new technologies might affect my personal privacy"
\begin{enumerate}
\item Strongly agree
\item Agree
\item Neutral
\item Disagree
\item Strongly disagree
\end{enumerate}

\item As far as you know, how much of what you do online or on your smartphone is being tracked by companies or organizations?
\begin{enumerate}
\item All or almost all of it
\item Most of it
\item Some of it
\item Very little of it
\item None of it
\end{enumerate}

\item As far as you know, how much of what you do online or on your smartphone is being tracked by the government?
\begin{enumerate}
\item All or almost all of it
\item Most of it
\item Some of it
\item Very little of it
\item None of it
\end{enumerate}

\item Thinking about the information available about you online, please tell me whether you’re concerned about social media companies gathering your personal information from their platform to create a profile of your interests and personal traits for marketing purposes
\begin{enumerate}
\item Extremely concerned
\item Concerned
\item Somewhat concerned
\item Not concerned
\end{enumerate}

\item Thinking about the information available about you online, please tell me whether you’re concerned about companies or organizations using information available about you online to make decisions about you, such as for a job, an insurance claim or health coverage
\begin{enumerate}
\item Extremely concerned
\item Concerned
\item Somewhat concerned
\item Not concerned
\end{enumerate}

\item Have you adjusted privacy settings on a social media account?
\begin{enumerate}
\item Yes
\item No
\item Not sure
\end{enumerate}

\item Have you deleted or stopped using a social media account because of privacy concerns?
\begin{enumerate}
\item Yes
\item No
\item Not sure
\end{enumerate}

\item Have you refused to provide an organization or business with your personal information because of privacy concerns?
\begin{enumerate}
\item Yes
\item No
\item Not sure
\end{enumerate}

\item Have you stopped doing business with a company that experienced a privacy breach?
\begin{enumerate}
\item Yes
\item No
\item Not sure
\end{enumerate}

\item Have you raised a privacy concern with a company or organization?
\begin{enumerate}
\item Yes
\item No
\item Not sure
\end{enumerate}

\item How often, if at all, do you read privacy policies, notices or pop-ups when using mobile applications or conducting transactions online?
\begin{enumerate}
\item Always
\item Sometimes
\item Never
\end{enumerate}

\item What is your primary reason for not always reading privacy notices? \textit{Please specify if choosing "Other"}
\begin{enumerate}
\item They are too long
\item They contain too much legal jargon
\item You don’t care
\item Other
\end{enumerate}

\item Have you or someone you know been impacted by a privacy breach?
\begin{enumerate}
\item Yes
\item No
\item I don't know
\end{enumerate}

\item There are a growing number of news reports of sensitive personal information being lost, stolen or made public. Has this had a major effect, moderate effect, minor effect or no effect at all on your willingness to share personal information with organizations?
\begin{enumerate}
\item Major effect
\item Moderate effect
\item Minor effect
\item No effect at all
\end{enumerate}

\item Why do you want to participate in this study?

\item Please provide us with a Canadian mailing address. \textit{We need this information to validate that you are a resident of Canada and to send you the gift card.}
\end{enumerate}
\section{Mapping to Participant's Background Attributes}
\label{a:mapping}

We used the following rules to assign a level to the background attributes of participants in the directed phase. The rules are based on the participants' answers to the questions Appendix~\ref{a:s2Questionnaire}.

\subsection*{Privacy Attitude}

A participant is assigned the level \textbf{High} if they answered \textit{Concerned} or \textit{Extremely Concerned} to Question 11, 15, and 16 and answered \textit{Yes} to at least two of the three questions: 17, 18, 19. A participant is assigned the level \textbf{Low} if they answered \textit{Not Concerned} to Question 11. A participant is assigned the level \textbf{Low} if they answered \textit{Somewhat Concerned} to Question 11 but answered \textit{No} to two of the three questions 17, 18, 19. Otherwise, a participant is assigned the level \textbf{Moderate}.

\subsection*{Digital Literacy}

A participant is assigned the level \textbf{High} if they answered (c) or (d) to Question 9. They are assigned the level \textbf{Moderate} if they answered (b). They are assigned the level \textbf{Low} if they answered (a).

\subsection*{Threat Awareness}

A participant is assigned the level \textbf{High} if they answered \textit{Strongly Agree} to Question 12 or if they answered \textit{Agree} to Question 12 and \textit{Very Closely} to Question 10. A participant is assigned the level \textbf{Low} if they answered \textit{Strongly Disagree}, \textit{Disagree}, or \textit{Neutral} to Question 10. Otherwise, a participant is assigned the level \textbf{Moderate}.

\end{document}